\documentclass{emulateapj}
\usepackage{amsfonts}

\shorttitle{Radio Loudness Distribution Dichotomy in Quasars}
\shortauthors{M. Balokovi\'{c} et al.}

\slugcomment{  }

\begin{document}

\title{Disclosing the Radio Loudness Distribution Dichotomy in Quasars: An Unbiased Monte Carlo Approach Applied to the SDSS-FIRST Quasar Sample}

\author{M.~Balokovi\'{c}\altaffilmark{1,2},
        V.~Smol\v{c}i\'{c}\altaffilmark{2,3,4},
        \v{Z}.~Ivezi\'{c}\altaffilmark{5},
        G.~Zamorani\altaffilmark{6},
        E.~Schinnerer\altaffilmark{7},
        B.~Kelly\altaffilmark{8}
        }

\altaffiltext{1}{California Institute of Technology, Astronomy Department, MC 249-17, 1200 East California Boulevard, Pasadena, CA 91125, USA}
\altaffiltext{2}{University of Zagreb, Physics Department, Bijeni\v{c}ka cesta 32, 10002 Zagreb, Croatia}
\altaffiltext{3}{Argelander-Institut f\"{u}r Astronomie, Auf dem Hugel 71, D-53121 Bonn, Germany}
\altaffiltext{4}{ALMA-COFUND Fellow}
\altaffiltext{5}{University of Washington, Department of Astronomy, Box 351580, Seattle, WA 98195, USA}
\altaffiltext{6}{INAF - Osservatorio Astronomico di Bologna, via Ranzani 1, I-40127 Bologna, Italy}
\altaffiltext{7}{Max-Planck-Institut f\"{u}r Astronomie, K\"{o}nigstuhl 17, D-69117 Heidelberg, Germany}
\altaffiltext{8}{University of California Santa Barbara, Department of Physics, Broida Hall, CA 93106, USA}

\begin{abstract}
We investigate the dichotomy in the radio loudness distribution of quasars by modelling their radio emission and various selection effects using a Monte Carlo approach. The existence of two physically distinct quasar populations, the radio-loud and radio-quiet quasars, is controversial and over the last decade a bimodal distribution of radio loudness of quasars has been both affirmed and disputed. We model the quasar radio luminosity distribution with simple unimodal and bimodal distribution functions. The resulting simulated samples are compared to a fiducial sample of 8,300 quasars drawn from the SDSS DR7 Quasar Catalog and combined with radio observations from the FIRST survey. Our results indicate that the SDSS-FIRST sample is best described by a radio loudness distribution which consists of two components, with $12\pm1$\% of sources in the radio-loud component. On the other hand, the evidence for a local minimum in the loudness distribution (bimodality) is not strong and we find that previous claims for its existence were probably affected by the incompleteness of the FIRST survey close to its faint limit. We also investigate the redshift and luminosity dependence of the radio loudness distribution and find tentative evidence that at high redshift radio-loud quasars were rarer, on average ’louder’, and exhibited a smaller range in radio loudness. In agreement with other recent work, we conclude that the SDSS-FIRST sample strongly suggests that the radio loudness distribution of quasars is not a universal function, and that more complex models than presented here are needed to fully explain available observations.
\end{abstract}

\keywords{Galaxies: surveys -- Cosmology: observations -- Radio continuum: galaxies  }

\section{Introduction}
\label{sec:intro}

An important property of Type~1 AGNs (i.e.\ broad line quasi stellar objects; QSOs) is the existence of radio-loud (RL) and radio-quiet (RQ) populations. One of the more controversial topics in QSO studies is whether these RL and RQ quasars\footnote{Throughout this paper we will use terms 'quasar' and 'QSO' interchangeably, referring to both quasi-stellar radio sources and quasi-stellar objects.} form two {\em physically distinct} populations of objects. On the one hand, both types of quasars are likely powered by similar physical mechanisms (e.g.\ Barthel 1989; Urry \& Padovani 1995; \citealt{shankar10}), and their radio loudness has been shown to anti-correlate with accretion rates (in Eddington luminosity units) onto their central supermassive black holes (e.g.\ \citealt{sikora07}). On the other hand, it has been demonstrated that, relative to RQ quasars, RL quasars are likely to reside in more massive host galaxies (e.g.\ Peacock, Miller, \& Longair 1986; \citealt{sikora07}), and harbor more massive central black holes (e.g.\ Laor 2000; Lacy et al. 2001; McLure \& Jarvis 2004). However, mainly due to two problems, i) severe selection biases, such as incompleteness in observed QSOs samples, and ii) the overwhelmingly high fraction of RQ quasars, it is still unclear whether RL and RQ quasars form two distinct populations of objects, or a continuous sequence (as suggested by e.g.\ \citealt{lacy01}). Here we focus on this problem.

Whether RL and RQ quasars form two distinct populations can be studied by investigating the relation between their radio and optical emissions. This illuminates the relative importance of the likely dominant sources of electromagnetic radiation in these two wavelength windows, namely the relativistic jet and the accretion disk. \citet{strittmatter80} first pointed out that the radio-to-optical flux density ratio for optically selected QSOs appears bimodal (the so-called 'quasar radio dichotomy'), which suggests that QSOs can be divided into two distinct, RL and RQ, populations of objects. Other authors found additional evidence for a dichotomy (e.g.\ \citealt{kellermann89,miller90,ivezic02,white07,zamfir08}). However, several studies (e.g.\ \citealt{white00,cirasuolo03b,lacy01}) disputed its existence. A bimodal distribution in quasar radio loudness would point to distinct physical properties of radio-loud and -quiet QSOs, such as a different physical origin of radio emission (jet vs.\ corona; \citealt{laor08}), different black hole masses \citep[e.g.][]{mclure04}, accretion rates (e.g.\ \citealt{sikora07,hamilton10}) and/or spins (e.g.\ \citealt{blandford77,blandford90,garofalo10}), as well as host galaxy properties (e.g.\ \citealt{sikora07,lagos09,kimball11}). The existence of two distinct quasar populations may then be linked to hierarchical structure growth in a $\Lambda$CDM universe in which dark matter halos and galaxy mergers play an important role \citep[e.g.][]{hughes03,lagos09,garofalo10,hamilton10,fanidakis10}.

The radio-to-optical flux density ratio as a measure of radio loudness was initially proposed by Schmidt (1970). A division between RL and RQ quasars at a ratio of $\sim$10 was found by Kellermann et~al.\ (1986) using rest-frame 5~GHz (6~cm) radio and B-band optical flux densities. An alternative radio loudness definition\footnote{For more details on definitions of radio loudness the reader is referred to Appendix~C in Ivezi\'{c} et al. (2002).} is purely based on radio luminosity, e.g.\ $L_{\mbox{\tiny{1.4~GHz}}}\gtrsim10^{25}$ W Hz$^{-1}$~\citep{miller90}. Studies of large QSO samples have yielded that the most RL QSOs are $\sim10^3$ times more luminous in radio than in optical (e.g.\ \citealt{ivezic02}). On the other hand, much deeper radio studies of smaller samples showed that RQ QSOs typically have a factor of $\sim10$ weaker radio than optical emission (e.g.\ \citealt{kellermann89}). Using SDSS \citep{york00} and FIRST \citep{becker95} data, \citet{ivezic04} have shown the existence of a peak at the high end of the radio-to-optical flux ratio distribution. However, because of the detection limits of both surveys, they could not detect a significant number of RQ objects at radio wavelengths, and hence argued the existence of a secondary peak at the low end (and thus bimodality) based on two arguments: i) the majority of SDSS quasars ($\sim90\%$) were not detected in the FIRST survey, thus they should lie at the low end of the radio loudness distribution, and ii) deep radio studies showed that typical RQ QSOs have a factor of $\sim10^4$ weaker radio emission than typical RL QSOs, thus a secondary peak must exist.

One of the most recent results on the quasar radio dichotomy comes from a stacking analysis of FIRST 20~cm snapshot images at the positions of $\sim40,000$ quasars from the SDSS DR3 catalog \citep{white07}. They showed that a shallow minimum exists between the RL and RQ parts of the radio loudness distribution; however, it was stressed that optical selection effects probably dominate the observed distribution of radio loudness. Specifically, the authors discuss the difficulty of identification of SDSS quasars at $2.4<z<3$, where their colors are very similar to stellar colors (see e.g., \citealt{richards02}), and support their claim by noting the strengthening of the bimodality for a subsample of quasars in that redshift range.

\citet{cirasuolo03b} have used Monte Carlo simulations of the quasar population in order to model the intrinsic radio loudness distribution. They have compared their simulated samples to three quasar samples (2dF QRS, LBQS and PBQS; see \citealt{cirasuolo03b} for details) with FIRST survey data and radio observations by \citet{kellermann89}. The three samples probe a wide range of radio loudness, but contain fewer than 200 radio-detected quasars in total. The authors reported that the best-fit radio-to-optical flux ratio distribution is a double-Gaussian function with $\sim$97\% of the quasars in the RQ, and $\sim$3\% in the RL component. However, they conclude that there is no minimum between the two Gaussians, thus no bimodality exists (but see \citealt{ivezic04} for a different interpretation). 

The main problems causing the discrepancies in literature regarding the existence or absence of a quasar radio loudness dichotomy lie either in the ambiguity of quasar selection, low number statistics, or severe selection biases linked to flux-limited samples (because radio loudness is defined as a radio-to-optical flux ratio, i.e.\ a ratio of two quantities drawn from flux limited samples). To properly address all of these biases, in the work presented here we utilize a Monte Carlo based approach similar to that in \citealt{cirasuolo03b}, but more robust and using a much larger sample of observed quasars (the largest QSO database currently available is the SDSS quasar catalog).

The outline of the paper is as follows. In Section~\ref{sec:data}\ we present the data and our sample. In Section~\ref{sec:algorithm}\ we outline our methodology and the simulation algorithms. We present our results in Section~\ref{sec:results}, and discuss them in Section~\ref{sec:discussion}. We summarize our conclusions in Section~\ref{sec:summary}. Throughout the paper we report magnitudes in the AB system, for the optical as well as for the radio. We use standard cosmology ($H_0=70$ km s$^{-1}$ Mpc$^{-1}$, $\Omega_M=0.3$, $\Omega_\Lambda=0.7$) and quasar continuum spectrum defined as $f_{\nu} \varpropto \nu^{\alpha}$, where $\nu$ denotes frequency, $f_{\nu}$ is flux density and $\alpha$ is the spectral index.

\section {Data and Samples}
\label{sec:data}

\subsection{Choice of the Source Catalog}

For a robust study of the radio loudness of quasars it is necessary to use a large, statistically significant radio-optical sample of quasars that covers the broadest possible range in radio loudness. As quasars are rare objects (e.g.\ the space density of SDSS quasars with $i<19$ is $\sim$11~deg.$^{-2}$) a statistical sample can be assembled via large-area observations. On the other hand, achieving radio sensitivities over these fields deep enough to probe far into the radio-quiet regime with the current generation of radio interferometers is challenging, and unfeasible over fields larger than a few degrees.

Because of the scaling of both radio and optical depth limits with surveyed areas, state-of-the-art surveys, such as e.g.\ SDSS-FIRST ($\sim$9380~deg.$^2$, $i<22.5$, $S_{1.4\mathrm{GHz}}>1$~mJy; \citealt{schneider10,becker95}), Stripe 82 (92~deg.$^2$, $g<24.5$, $S_{1.4\mathrm{GHz}}>52~\mu$Jy; \citealt{hodge11}), COSMOS (2~deg.$^2$, $i<26.5$, $S_{1.4\mathrm{GHz}}>50~\mu$Jy; Scoville et al.\ 2007; Schinnerer et al.\ 2007), and VVDS (1~deg.$^2$, $I<24$, $S_{1.4\mathrm{GHz}}>80~\mu$Jy; \citealt{lefevre03,bondi03}) cover a comparable radio loudness range, while smaller area surveys suffer from small number statistics in their optical quasar samples (see e.g.\ Smol\v ci\' c et al.\ 2008). Hence, for the analysis presented here we have utilized the largest available sample of quasars with radio coverage, drawn from the SDSS and FIRST sky surveys. 

For the reasons outlined above, we use the SDSS Seventh Data Release Quasar Catalog \citep{schneider10}. It contains 105,783 spectroscopically confirmed quasars taken from $\sim$9,380~deg.$^2$ of the sky. The catalog consists of objects with spectroscopy that 1) have reliable redshifts, 2) have at least one emission line with FWHM greater than 1000~km~s$^{-1}$\ or interesting/complex absorption features, and 3) are more luminous than rest-frame $M_i=-22.0$ (in our cosmological model, assuming a power-law continuum with $\alpha=-0.5$). The selection is based on sources' $i$ magnitude and position in the multidimensional color-space based on five SDSS photometric bands (see \citealt{richards02} for details). The quasars in the sample have $15.0<i<21.2$, but the majority of quasars are brighter than $i\approx19$ because the flux limit for the main spectroscopic sample is $i<19.1$. In addition to the main spectroscopic sample, the catalog contains serendipitously identified quasars and sources selected for their proximity (within 2'') to a 1.4~GHz radio source drawn from the FIRST survey. The rest of the quasars in the sample are matched to the FIRST catalog to within a 2'' radius. The radio flux densities at 20~cm (1.4 GHz) are given in "AB radio magnitudes":
\begin{equation}
t = -2.5 \log \left( \frac{f_{\mbox{\tiny{rad}}}}{3631 \mbox{Jy}} \right) \mbox{,}
\end{equation}
where $f_{\mbox{\tiny{rad}}}$ is the radio flux density at 20~cm in Jy.

FIRST is a radio survey at 1.4~GHz/20~cm, conducted with the Karl G. Jansky Very Large Array' (VLA) in B-configuration \citep{becker95}. It has mapped approximately 10,000~deg.$^2$ of the North Galactic Cap with a beam size of 5.4'' and a typical RMS sensitivity of 0.15 mJy/beam. Flux density values used here are integrated over the two-dimensional Gaussians fitted to each source. Due to the lack of very short spacings in the VLA B-configuration and the nature of the Gaussian-fitting source detection algorithm, fluxes for extended objects larger than about 10'' are likely underestimated and split into multiple components in the FIRST source catalog. We estimate that this does not significantly affect our sample as most of the radio-detected quasars are expected to be unresolved at the angular resolution of the FIRST survey. Multiple component sources are rare and in most cases radio-loud, which puts them high above the interesting transition region between RQ and RL regimes (e.g.\ Jiang et al. 2007). For a discussion of the distribution of integrated and peak flux densities and the extended source bias of FIRST-detected SDSS quasars we refer the reader to \citet{kimballivezic08}. Nonetheless, we do take this effect into account statistically, using the FIRST survey completeness correction derived specifically for SDSS quasars (see Figure 1 in Jiang et al. 2007, and references therein), taking into account both the source size and flux distribution.

\subsection{Our Main Sample}
\label{sec:mainsample}

Following the SDSS DR7 Quasar Sample notation, hereafter we adopt the apparent radio magnitude ($t$) for radio flux densities. The catalog lists 8,630 quasars with a radio detection within 2'' from the optical source position. For 'radio luminosity' here we use the 1.4~GHz/20~cm absolute radio magnitude ($M_t$), derived from the apparent radio magnitude using a K-correction of the form $-2.5(1+\alpha)\log(1+z)$ (e.g.\ Richards et~al.\ 2006), assuming $\alpha=-0.5$ for each quasar (e.g.\ Kimball \& Ivezi\'{c} 2008). For the 'optical luminosity' we use the emission-line-corrected absolute magnitude in the rest-frame SDSS $i$ band ($\lambda_{eff} = 7471$ \AA). The reason for this choice is that the SDSS $i$-band was used to construct the flux-limited DR7 quasar sample and that $i$ magnitudes are available even for quasars at the highest redshift. Prior to computing the absolute $i$ magnitudes for the quasar continua, we have corrected the cataloged apparent magnitudes for galactic extinction using the extinction maps of \citet{schlegel98}. To get continuum magnitudes, we subtract the contribution of emission lines to the $i$ band using a function derived by \citet{richards06}. Using the mean SDSS quasar spectrum, they have calculated the contributions of major quasar emission lines (above the power-law continuum) as a function of redshift, up to $z \approx 5$. After the subtraction of this contribution to the $i$-band apparent magnitudes, we use the canonical K-correction with $\alpha=-0.5$. Given the high uncertainty of this correction for $z \ge 5$, hereafter we exclude 56 quasars with $z \ge 5$ from our sample. For details of the calculation see Section 5 and Figure 15 in \citet{richards06}. In order to get a 'cleaner' optical selection, we exclude the quasars selected only on the basis of their proximity to a FIRST source. After the exclusion of those quasars, the $z>5$ quasars and those with uncertain photometry, we are left with 8,307 quasars in our main radio-optical subsample. Our main optical sample consists of 98,544 quasars.

\section{The Methodology for Monte Carlo Simulations}
\label{sec:algorithm}

If one had a complete volume-limited quasar sample with both rest-frame optical and radio luminosities, studies of the radio loudness distribution would be simple. For example, if radio and optical luminosities were linearly correlated (regardless of the physics of such a setup), the radio-to-optical luminosity ratio distribution would be very narrow and the radio versus optical luminosity diagram would show a straight narrow band. Another possible situation could be a broad distribution of the radio-to-optical luminosity ratio, with perhaps evidence for its two-component nature (dichotomy), with or without evidence for a local minimum (bimodality). In this case, the radio versus optical luminosity diagram would show a broadly dispersed line or two possibly overlapping lines, one for the RQ and one for the RL quasars (see the bottom two panels of Figure~\ref{fig:models} for an illustration). If the sampled redshift range was sufficiently broad, one could even perform such studies for subsamples selected from narrow redshift slices, and search for evidence of evolution in the inferred properties with cosmic time.

Unfortunately, such a complete sample does not exist. The main difficulty when working with real samples is that both optical and radio data are truncated at finite flux levels, and this strong selection effect must be properly accounted for. There are additional effects, such as K-corrections and optical variability, which can be handled to some extent, as discussed further below. In order to utilize existing samples, various statistical methods have been used in published work to account for the observational effects, as briefly reviewed in Section~1.

Here we use Monte Carlo simulations, with the following main logical steps:
\begin{enumerate}
\item The main sample is defined at optical wavelengths, with various selection effects, and the main observable is the optical apparent magnitude. The sample also contains redshifts for all quasars and radio magnitudes for a subset of radio-detected quasars. The selection function for the sample is well known.
\item Only a subset of real optical sources from the main sample are detected at radio wavelengths, comprising the main radio-optical subsample. They have a certain distribution of radio magnitudes and, at least in principle, a different distribution of optical magnitudes than the main optical sample. The distributions of radio and optical magnitudes for the radio-optical subsample are the main constraints on models for the relationship between the radio and optical luminosities.
\item Given a quasar from the main optical sample with certain optical luminosity, a parametrized model generates its radio luminosity (as detailed below). Using its implied apparent radio magnitude and the radio selection function, this object is either retained or rejected from the simulated radio-optical sample.
\item Starting with the main optical sample, the simulation produces a subset of quasars with simulated radio magnitudes -- a simulated radio-optical subsample. The distribution of simulated radio magnitudes for this subsample, as well as its corresponding distribution of optical magnitudes, is compared to observed radio and optical magnitude distributions of the main radio-optical subsample, and utilized in a $\chi^2$ minimisation procedure to get the best-fit model parameters.
\end{enumerate}

Although the Monte Carlo sample utilizes observed optical magnitudes, its radio magnitudes are generated stochastically. Therefore, there is no object-to-object correspondence between the real radio-optical subsample and the simulated radio-optical subsample. For a good model, we expect a statistical agreement for two main quantities: i) optical magnitude distribution, and ii) radio magnitude distribution between the observed and simulated radio-optical subsamples. For simulating the radio luminosities we initially use four models independent of redshift and optical luminosity and the complete main sample presented in Section~\ref{sec:mainsample}. However, once we select the best model, we investigate the redshift evolution and dependence on optical luminosity for its parameters. The following subsection defines radio loudness and describes the models considered in this work. The rest of this section describes the simulation algorithm (Section \ref{sec:simul}), the optimization strategy (Section \ref{sec:optimization}) and the evaluation of our method on purely artificial data (Section \ref{sec:artif}).

\subsection{The Radio Luminosity Models}
\label{sec:models}

We use four different models for the relationship between the radio continuum luminosity at 1.4~GHz ($L_{\mbox{\tiny{rad}}}$) and optical continuum luminosity in the SDSS~$i$ band ($L_{\mbox{\tiny{opt}}}$). Their relationship is parametrized using radio loudness ($R_K$) defined as:
\begin{equation}
R_K = \log \left( L_{\mbox{\tiny{rad}}} \right) - K \log \left( L_{\mbox{\tiny{opt}}} \right) \mbox{.}
\label{eq:Rdef}
\end{equation}
Here $K$ is either 0 or 1 and it selects one of the two different families of models that correspond to two common definitions of radio loudness found in the literature. For $K=0$ the distribution of radio loudness simply equals the distribution of radio luminosities (thus RL quasars are defined as being more luminous than some threshold radio luminosity; e.g.\ \citealt{peacock86,miller90}, also examined in \citealt{ivezic02} and \citealt{jiang07}), while for $K=1$ radio loudness is defined as the logarithm of the radio-to-optical luminosity ratio (with RL quasars having this ratio greater than some threshold; e.g.\ \citealt{kellermann89,ivezic02,ivezic04,cirasuolo03a,cirasuolo03b,jiang07}). Each definition is suitable for a specific model of the relationship between radio and optical luminosities.

\begin{figure}
\begin{center}
\includegraphics[scale=0.68]{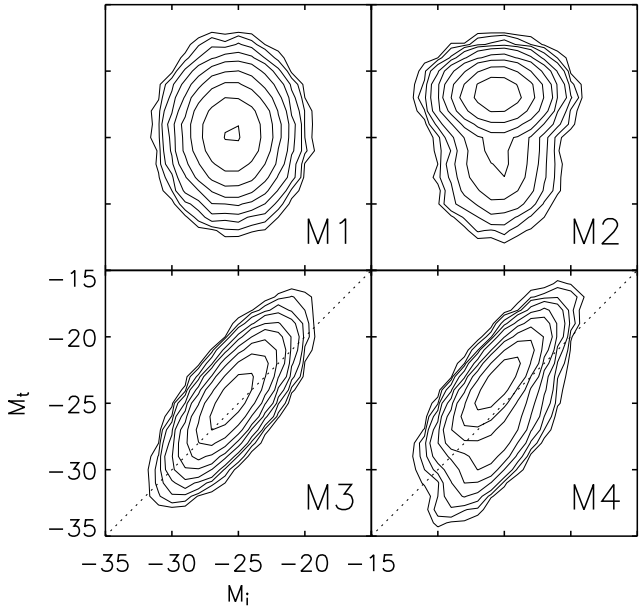}
\caption{Examples of the radio models (M1-4) considered in this paper plotted in the $M_i$-$M_t$ plane. The radio loudness distribution is a single Gaussian for models M1 and M3 and a double Gaussian for models M2 and M4. Note that, in general, the Gaussians forming the two-component radio loudness distribution may exhibit complete, partial or no overlap, regardless of the model. In models M1 and M2 the radio absolute magnitude ($M_t$) is independent of the optical absolute magnitude ($M_i$), while in models M3 and M4 $M_t$ is proportional to $M_i$ for every quasar. For clarity axis labels are shown only for the bottom-left panel, but they are the same for all panels. The dotted lines show where $M_t=M_i$. \label{fig:models}}
\end{center}
\end{figure}

We have considered four simple models for the radio luminosity: two in which it is independent of the optical luminosity and two in which it is directly proportional to the optical luminosity. In the former case radio loudness is properly defined with $K=0$, $R_0=\log \left( L_{\mbox{\tiny{rad}}} \right)$ and radio luminosity is modeled as $L_{\mbox{\tiny{rad}}} = 10^{R_0}$. In the latter case, the proper definition of radio loudness is $K=1$, $R_1=\log \left( L_{\mbox{\tiny{rad}}} / L_{\mbox{\tiny{opt}}} \right)$, hence the radio luminosity model is $L_{\mbox{\tiny{rad}}} = L_{\mbox{\tiny{opt}}}\times10^{R_1}$. In both cases, we examine single- and double-Gaussian distributions of radio loudness. The designations of the models in this paper and their basic descriptions (where PDF stands for probability density function) are:
\begin{itemize}
 \item M1: $K=0$, $R_0$ has a Gaussian PDF
 \item M2: $K=0$, $R_0$ has a double Gaussian PDF
 \item M3: $K=1$, $R_1$ has a Gaussian PDF
 \item M4: $K=1$, $R_1$ has a double Gaussian PDF
\end{itemize}
Examples of all four models are plotted in Figure~\ref{fig:models}. Models M1 and M3 are two-parameter models, with the free parameters being $x$ (the mean of the Gaussian) and $\sigma$ (the width of the Gaussian). Models M2 and M4 have five free parameters: two for each Gaussian ($x_1$, $\sigma_1$, $x_2$ and $\sigma_2$) and an additional parameter $f$, which determines the ratio between the integrals of the two Gaussians. In all cases the overall normalization is determined automatically by the requirement that the number of radio-detected quasars in the simulated radio-optical samples matches the number in the observed SDSS-FIRST sample.

Initially, we use these four models to simulate radio-optical samples based on the complete main optical sample of 98,544 SDSS quasars and match them to the radio-optical subsample. Later we search for redshift evolution and dependence of the radio loudness distribution shape on optical luminosity not by adjusting the models, but rather by separating the main sample into smaller samples constrained in redshift and optical luminosity. Possible trends can then be inferred by observing how best-fit model parameters change between different subsamples.

\subsection{The Simulation Algorithm}
\label{sec:simul}

\subsubsection{The Main Optical Sample}
\label{sec:simulation1}

For all radio loudness models that we have considered, the simulated radio magnitude~($t$) was calculated for each optically detected quasar in the main optical sample presented in Section~\ref{sec:mainsample}. This sample contains 98,544 optically detected quasars with all of the selection biases introduced by the DR7 Quasar Catalog, except for the radio-only selection which was eliminated from the catalog for this work. Since we compare our simulated radio-optical subsamples to the radio-optical subsample of that catalog (i.e.\ to the sources also detected in the radio), this biased sample is indeed the most valid optical sample for our simulations.

Another option for the optical sample would be a simulated sample produced from an empirical luminosity function. However, such an approach suffers from two main problems: i) a simulated optical sample would be free of selection biases (assuming selection bias was properly corrected in the construction of the luminosity function) and therefore we would additionally need to simulate the SDSS optical selection; and ii) observed optical counts are difficult to simulate with proper uncertainties included (e.g.\ from photometry uncertainty, conversion between photometric bands, K-corrections, optical variability etc.) unless the same uncertainties have been considered during the generation of the luminosity function (see e.g.\ \citealt{lafranca97} for an approach that does involve substantial consideration of uncertainties). 

\subsubsection{Assignment of Apparent Radio Magnitudes and Application of the Radio Selection Function}
\label{sec:simulation2}

In each simulation, we assign an apparent radio magnitude~($t$) to each quasar from the main optical sample. The apparent radio magnitude $t$ is calculated from the absolute radio magnitude $M_t$, which is modelled according to Equation~(\ref{eq:Rdef}) as
\begin{equation}
\label{eq:radioabsmag}
M_t = K M_i - 2.5 R_K,
\end{equation}
where $K$ takes on values of 0 or 1 and $R_K$ is a random variable drawn from either a Gaussian or a double-Gaussian probability distribution, depending on the model (see Section~\ref{sec:models}).

To convert from $M_t$ to $t$ we use a K-correction of the form $-2.5(1+\alpha)\log(1+z)$ (e.g.\ Richards et al.\ 2006), assuming $\alpha=-0.5$ for each quasar (e.g.\ Kimball \& Ivezi\'{c} 2008), and including corresponding uncertainties. Thus,
\begin{equation}
t = M_t + DM(z) + K_{\mbox{\tiny{c,rad}}}(z) + n(\sigma_t),
\label{eq:t}
\end{equation}
where $DM(z)$ is the distance modulus and $K_{\mbox{\tiny{c,rad}}}(z)$ is the K-correction for $\alpha=-0.5$, both dependent on quasar's redshift. The last term is a normally distributed random variable accounting for uncertainties in K-corrections and photometry. Its mean is zero and the standard deviation ($\sigma_t$) can be estimated by considering the scatter in the measured radio spectral indices (e.g.\ \citealt{kimballivezic08}). As the standard deviation in spectral indices is approximately 0.3, it follows from the definition of the K-correction that $\sigma_t \approx 0.35$ for $z=2$, where the number density of quasars is maximal. This uncertainty completely dominates the small photometric uncertainty (typically 0.03-0.04 mag).

For our $K=0$ models (M1 and M2) Equation~(\ref{eq:radioabsmag}) simplifies to $M_t = -2.5 R_0$, hence
\begin{equation}
\label{eq:tk0}
t = -2.5 R_0 + DM(z) + K_{\mbox{\tiny{c,rad}}}(z) + n(\sigma_t=0.35) \mbox{.}
\end{equation}
For our $K=1$ models (M3 and M4)
\begin{equation}
\label{eq:tk1temp}
t = M_i -2.5 R_1 + DM(z) + K_{\mbox{\tiny{c,rad}}}(z) + n(\sigma_t=0.35) \mbox{.}
\end{equation}
The absolute optical magnitude is
\begin{equation}
M_i = i - DM(z) - K_{\mbox{\tiny{c,opt}}}(z) + n(\sigma_i) + e(\sigma_v) \mbox{,}
\end{equation}
where $DM(z)$ is the distance modulus, $K_{\mbox{\tiny{c,opt}}}(z)$ is the K-correction and the remaining two terms account for the scatter in K-correction, photometry (normal distribution, $n(\sigma_i)$), and variability (exponential distribution, $e(\sigma_v)$). The optical spectral index distribution with an average of $-0.5$ and a standard deviation of 0.3 \citep{richards06} yields $\sigma_i \approx 0.35$, taking into account that the optical photometric errors (typically $0.03$ magnitude) are negligible. Optical variability scatter has an empirically determined exponential distribution with zero mean and $\sigma_v \approx 0.2$ \citep{ivezic04var}. Finally, inserting the expression for $M_i$ into Equation~(\ref{eq:tk1temp}), the distance moduli cancel out, as well as the mean K-corrections (because of our assumption of the same mean spectral index in the optical and radio bands), and the normally distributed uncertainties $\sigma_t$ and $\sigma_i$ add up in quadrature, yielding a convenient expression for $t$ in our $K=1$ models:
\begin{equation}
\label{eq:tk1}
t = i -2.5 R_1 + n(\sigma=0.5) + e(\sigma_v=0.2).
\end{equation}

Our simulation algorithm assigns radio magnitudes to optically detected quasars by drawing random numbers from appropriate distributions specified by Equations~(\ref{eq:tk0}), for M1 and M2, and~(\ref{eq:tk1}), for M3 and M4. Each simulated radio-optical sample (consisting of 98,544 quasars with apparent optical and radio magnitudes) is then subjected to the radio selection function in order to produce a smaller subsample of 'radio-detected' quasars corresponding to the observed radio-optical sample from SDSS and FIRST. All quasars fainter than the FIRST flux limit at 1~mJy are rejected. Some of the quasars above the FIRST flux limit are randomly rejected in order to mimic survey incompleteness. This is performed by randomly choosing and excluding a number of quasars in narrow $t$-magnitude bins, with the fraction of excluded objects being determined by the completeness function of the survey up to its flux limit (see Figure~1 in \citealt{jiang07}). At the end of this procedure, each simulated radio-optical subsample is one particular realization of the real radio-optical subsample drawn from SDSS and FIRST. The simulated radio-optical subsamples after radio selection have to consist of approximately 8,300 quasars to at least roughly match the observations, which sets the overall normalization for our models.

\subsubsection{Evaluation of Goodness of Fit for the Simulated Samples}
\label{sec:chi2evaluation}

Each of the simulated radio-selected subsamples ("realizations" hereafter) is binned into 10 bins in $t$, $i$, $z$ and $R'=0.4(i-t)$ distributions. We chose to examine these four distributions to provide better constraints for our models; in principle, weaker constraints, always consistent with those derived by using all four distributions, can be can be obtained with any subset of these distributions that includes either $t$ or $R'$. Note that the $R'$ distribution is a proxi to the radio loudness distribution in the case of $K=1$; otherwise it is just an additional constraint on the relation between $t$ and $i$ magnitudes\footnote{Also note that this distribution is severely biased and cannot be directly used to infer the distribution of radio loudness; see Ivezi\'{c} et al. (2002, 2004) for an explanation.}. The bin values are added to a pool of realizations simulated with identical model parameters. The radio magnitude assignment, the selection procedure and the binning need to be repeated a large number of times for a given model and a set of its parameters in order to properly account for the stochasticity of particular realizations. The aim of this Monte Carlo procedure is to calculate the mean distributions of $t$, $i$, $z$ and $R'$ for a given set of model parameters and derive their expected variance. We calculate the mean and the standard deviation of the number of simulated objects in each bin ($SN_j$ and $\sigma_j$ in the equation below) from a large number of realizations and compute the total $\chi^2$ with respect to the binned $t$, $i$, $z$ and $R'$ distributions of the real SDSS-FIRST data. The $\chi^2$ is defined as
\begin{equation}
\chi^2 = \sum_{j=1}^{N} { \left( \frac{RN_j-SN_j}{\sigma_j} \right) }^2 \mbox{,}
\label{eq:chisqared}
\end{equation}
where the sum runs over all 40 bins, while $RN_j$ is the number of real quasars in a particular bin and $SN_j$ and $\sigma_j$ are the mean and the standard deviation for that bin, determined from the simulations as described above.

Although the four binned distributions are not entirely independent of each other (in particular, $R'$ is a linear combination of $i$ and $t$), the tests on artificial data (see Section~\ref{sec:artif}) have shown that the number of degrees of freedom may be assumed as if the bins were independent. Therefore, the number of degrees of freedom equals $N-p-1$, where $N$ is the total number of bins used for evaluation and $p$ is the number of free parameters of a particular model. Hereafter we refer to one evaluation of $\chi^2$ from its distribution given a large number of realizations, as a "simulation" and abbreviate the number of realizations per simulation with $N_r$.

\begin{figure}
\begin{center}
\includegraphics[scale=0.5]{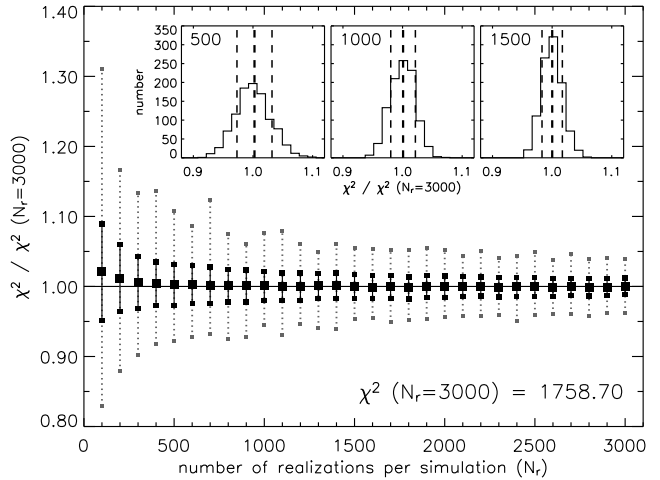}
\caption{ An illustration of convergence of $\chi^2$ to a unique value (represented here as the mean $\chi^2$ for $N_r=3000$) as the number of realizations per simulation ($N_r$) increases. While smaller $N_r$ requires less computing time, such simulations have greater variance of $\chi^2$ between them. For each given number of realizations per simulation, the same model (M4 here, with all parameters fixed) was simulated 1000 times in order to find the mean (large black rectangles), the standard deviation (smaller black rectangles at the ends of black lines) and minimum/maximum values of $\chi^2$ (gray rectangles at the ends of gray dotted lines) for a given $N_r$. The insets show distributions of $\chi^2$ values for three selected $N_r$ values indicated in the upper left corner of each inset. The dashed vertical lines in the insets show the mean (thick) and the standard deviation (thin lines). \label{fig:nstack}}
\end{center}
\end{figure}

It is a general property of Monte Carlo simulations to converge to an average result only after the experiment is repeated a large number of times. We test the convergence by examining how the means $\chi^2$, standard deviations and minimum/maximum values for different simulations with same $N_r$. This is illustrated in Figure~\ref{fig:nstack} for model M4 (analogous results follow from any of the models considered here) with a fixed arbitrarily chosen set of its parameters and a range of different $N_r$. Clearly, a larger number of realizations per simulation reduces the probability of getting outlying $\chi^2$ values that significantly deviate from the mean $\chi^2$ for the simulated model, but at the expense of computing time. We have found it more time-effective to work with a relatively small number of realizations in our optimization procedure and correct for the scatter in the $\chi^2$ values just before deriving the marginal probability density functions for model parameters (see Section \ref{sec:optimization}). Due to computing time constraints, we chose to calculate 1000 realizations per simulation as a reasonable compromise between accuracy and computing time. The calibration shown in Figure \ref{fig:nstack} shows that for $N_r = 1000$ one should expect variations with standard deviation of $\sim$2\%. For $N_r = 3000$ the variation drops to 1.2\%, which does not represent a significant improvement compared to $N_r = 1000$. The computing time scales linearly with the number of realizations per simulation, so $N_r = 3000$ would require three times more computing time than $N_r = 1000$.

\subsection{The Optimization Strategy}
\label{sec:optimization}

Since there is no a priori indication that the parameter space for our models is simple, e.g.\ that there is a single minimum of the total $\chi^2$, we began to search the parameter space with a random walk across a wide range of parameters. The algorithm we used to initially sample the parameter spaces with 2 to 5 dimensions, is a Metropolis algorithm (Metropolis et al. 1953). Several runs of the algorithm were used to identify the area around the global minimum of $\chi^2$ for each of the models. The program would then proceed to evaluate the region on a regular grid, so that marginal probability distributions could be estimated. We have used three levels of refinement of the grid in order to be able to sample well the narrow region around the global minimum of $\chi^2$.

Since variations displayed in Figure \ref{fig:nstack} are expected to occur in the evaluation process, we needed to take into account the fact that some simulations resulted in $\chi^2$ values as low as the minimum $\chi^2$ value, although their mean $\chi^2$ would be much larger. For all parameter space points for which $\chi^2$ was found to be less than 5 $\sigma_{\chi^2}$ (recall that this is $\sim$2\% for $N_r = 1000$) above the lowest value found, the optimization would proceed to map the surrounding parameter space in more detail (typically, with a factor of 3 more resolution in all dimensions).

For computation of the marginal probability density functions for model parameters, we assigned them the probability normally associated with $\chi^2$ values:
\begin{equation}
p (\vec{x}) \propto e^{-\frac{1}{2}\chi^2 (\vec{x})},
\label{eq:probability}
\end{equation}
where $\vec{x}$ is a vector (a set of coordinates) in the parameter space of some model. The marginal probability of a certain parameter value is then the sum of the probabilities over all other parameter space dimensions. The probability density functions were normalized a posteriori so that the sum of probabilities of each bin in the parameter values equals unity. To cope with the problem of the internal scatter in $\chi^2$ values (see Section \ref{sec:chi2evaluation}), we used a special procedure to compute realistic probability density functions which take into account that the computed $\chi^2$ may vary according to Figure \ref{fig:nstack}. First, we estimate the standard deviation of $\chi^2$ from the calculations illustrated in Figure \ref{fig:nstack} and described in the caption; e.g.\ the fractional standard deviation for $N_r = 1000$ is $\sim$2\%. Then, we calculate the corrected marginal probability density functions by adding a normal random variate with standard deviation equal to 2\% of the lowest $\chi^2$ value to the previously computed $\chi^2$ values, calculating the probability distributions in each case and repeating this procedure many times. This Monte Carlo method requires $\sim$1000 repetitions for the resulting mean marginal probability density function to converge. We derive the asymmetric error bars on parameter values as intervals containing $\pm34$\% of the total probability around the median of the marginal probability density function for each parameter.

\subsection{Verification of the Approach on Artificial Datasets}
\label{sec:artif}

In order to evaluate our method and better understand the possible uncertainties and biases it involves, we first simulated artificial datasets resembling real data in all aspects, differing only in the fact that their radio loudness distributions are exactly known. Treating them the exact same way as the observational data, we have reconstructed their model parameter values and compared them to the input ones.

\begin{figure}
\begin{center}
\includegraphics[scale=0.52]{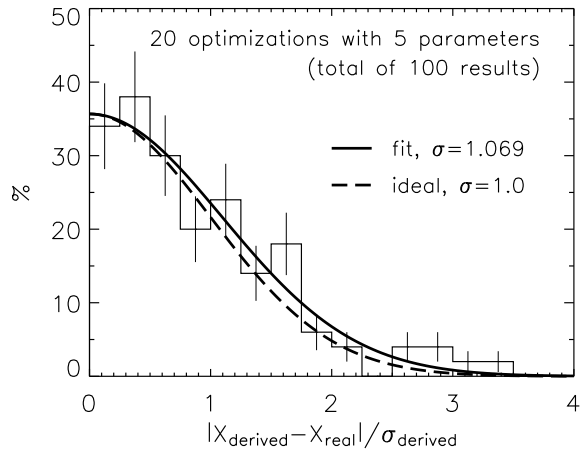}
\caption{ The distribution of parameters derived from our parameter optimization procedure (X$_{\mbox{derived}}$) relative to the input parameters of our artificial datasets (X$_{\mbox{real}}$), normalized to the derived standard deviation of the reconstructed parameters ($\sigma_{\mbox{derived}}$). The data is displayed for 20 optimizations on artificial samples constructed with model M4, which has 5 free parameters. The vertical error bars represent Poissonian noise. The solid black line marks the Gaussian fitted to the data, which is a close match to the ideal Gaussian with $\sigma=1$ (dashed line), verifying that with our method it is possible to reconstruct the model parameters as expected from statistics. \label{fig:testing}}
\end{center}
\end{figure}

For a set of artificial datasets constructed with the radio models presented in Section~\ref{sec:models}, the parameters were reconstructed to within 3$\sigma$ uncertainties in all trials. The distribution of the reconstructed parameters, normalized to their respective standard deviations, is displayed in Figure~\ref{fig:testing}. The lowest $\chi^2$ values reached in optimizations were $\chi^2 \approx 35$, yielding a reduced $\chi^2$ of~$\sim$1 if the problem has the number of degrees of freedom as if the fitted distributions were independent (see Section~\ref{sec:chi2evaluation}). 

We have also tried fitting an M4 ($K=1$) model to artificial datasets constructed with model M2 ($K=0$). A good fit should not be possible in that case. The lowest $\chi^2$ values in most of these cases stayed above $\sim$1000. The fitted parameters present in both models (e.g.\ $x_1$, $x_2$) would often be several standard deviations from their correct values. We consider this result to be an indication of what may be expected if none of the models considered in this work yield an adequate description of the observed data. Tests performed with other models are consistent with the above results.

\section{Results}
\label{sec:results}

\subsection{Models Fitted to the Complete Main Sample}

\begin{table*}[!ht]
\caption{Best-fit median values and 1$\sigma$ uncertainties of model parameters for our four models, obtained from fits to the complete main sample of quasars. Value of $K$ is a propery of the models and parameters describe the Gaussian peaks ($x_1$ and $x_2$), widths ($\sigma_1$ and $\sigma_2$), and their relative normalization ($f$). See Section~\ref{sec:models} for details. \label{tab:results-all}}
\begin{center}
\begin{tabular}{|c|c|c|c|c|c|c|c|}
\hline
model & \ $\chi_{\mbox{\tiny{MIN}}}^2$ \ & $K$ & $x_1$ & $\sigma_1$ & $f$ & $x_2$ & $\sigma_2$ \\
\hline \hline
M1 & 4032 & 0 & $ 8.4_{-0.1}^{+0.1} $ & $ 2.0_{-0.1}^{+0.1} $ & 0 & 0 & 0 \\
M2 & 2450 & 0 & $ 9.53_{-0.04}^{+0.02} $ & $ 0.1_{-0.05}^{+0.08} $ & $ 0.11_{-0.01}^{+0.01} $ & $ 11.8_{-0.1}^{+0.1} $ & $ 0.75_{-0.07}^{+0.09} $ \\
M3 & 2034 & 1 & $ -1.4_{-0.1}^{+0.1} $ & $ 1.9_{-0.1}^{+0.1} $ & 0 & 0 & 0 \\
M4 & 1053 & 1 & $ -0.11_{-0.04}^{+0.07} $ & $ 0.42_{-0.07}^{+0.06} $ & $ 0.12_{-0.02}^{+0.02} $ & $ 1.4_{-0.2}^{+0.2} $ & $ 1.03_{-0.06}^{+0.09} $ \\
\hline
\end{tabular}
\end{center}
\end{table*}

\begin{figure*}
\includegraphics[width=\textwidth]{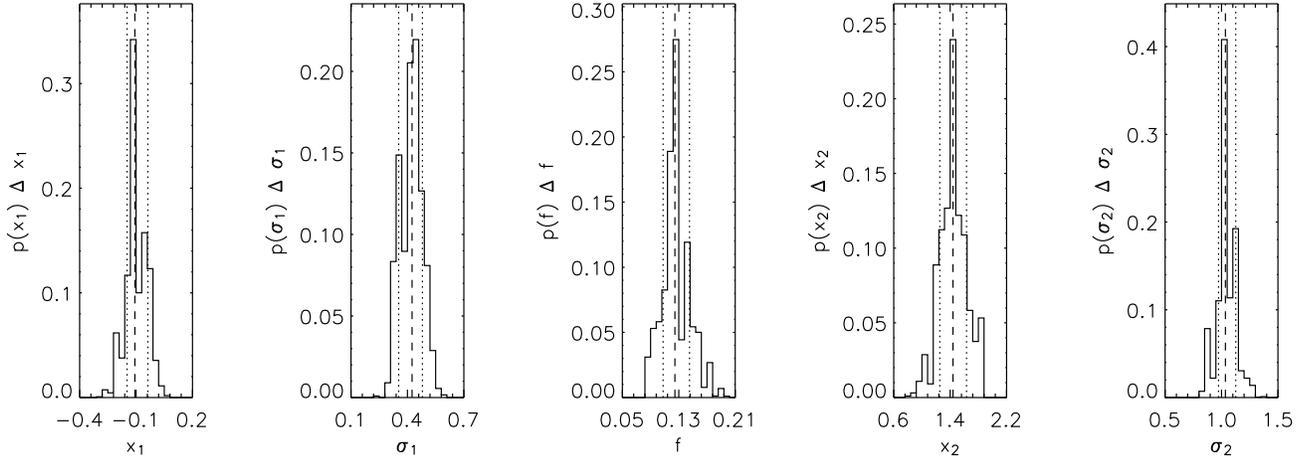}
\caption{ The marginal probability distributions for all 5 parameters of our model M4. The parameters $x_1$ and $\sigma_1$ set the radio-quiet portion of the radio loudness distribution, $f$ is the fraction of radio-loud quasars and $x_2$ and $\sigma_2$ set the radio-loud part. The distributions were computed through Monte Carlo simulations that compensate for the instability of the $\chi^2$ computed using a low number of realizations per simulation (see Section~\ref{sec:optimization} for details). \label{fig:mcmarginal-M4}}
\vspace{0.3cm}
\end{figure*}

We attempt to reproduce the observed SDSS-FIRST sample by performing simulations to the whole main sample described in Section~\ref{sec:mainsample}. The results of these optimizations are given in Table~\ref{tab:results-all} for all four models. Based on our analysis, the least likely models to fit the observations are M1 and M2, in which the radio luminosity of quasars is independent of their optical luminosity ($K=0$). The minimum total $\chi^2$ values reached in our optimization procedure were $\sim$4000 and $\sim$2500 for models M1 and M2, respectively. Each of the models has 37 degrees of freedom. For both models, the largest contribution to the total $\chi^2$ comes from the disagreement with the observed redshift distribution. They also fail to correctly reproduce the most populated bins in the radio and optical magnitude distributions. Model M2, however, fits the bright end of the radio magnitude distribution considerably better and this is clearly reflected in the lower $\chi^2$ value. Values of $\chi^2$ generally fluctuate with $\sigma_{\chi^2}<100$ (see Figure \ref{fig:nstack} and Section \ref{sec:chi2evaluation}), making the difference of 1500 in total $\chi^2$ of the two models highly significant.

Models M3 and M4, both involving a proportionality between the optical and the radio luminosity of quasars ($K=1$), were found to provide better fits to the main sample compared to models M1 and M2. Their lowest $\chi^2$ values of $\sim$2000 and $\sim$1000, respectively, are significantly lower than for models M1 and M2. Both models can reproduce the observed redshift distribution much more correctly (especially the lowest redshift bin), but the overall match is not a statistically good one since both have only 34 degrees of freedom. The most significant contributors to the total $\chi^2$ are the optical magnitude bins with the highest numbers of quasars, with the bright end of the radio magnitude distribution contributing considerably to the higher $\chi^2$ value for model M3. The non-optimal matching in the radio and optical magnitude distributions results in discrepancies in the $R'=0.4(i-t)$ distribution as well.

\subsection{The Best-fit Model for the Complete Main Sample}

\begin{figure}
\begin{center}
\includegraphics[scale=0.5]{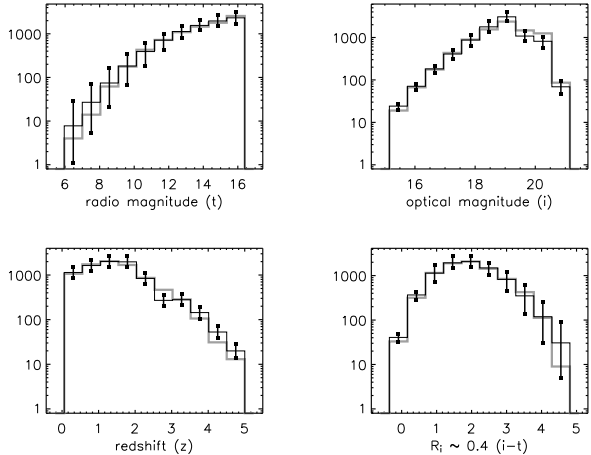}
\caption{ Distributions of $t$ and $i$ magnitudes, redshift and $R'=0.4(i-t)$ for the observed SDSS-FIRST sample (thick gray histograms) and for a simulation with the best-fit model M4 and parameter values set to the median values of their respective marginal probability distributions (thin black histograms; see Table~\ref{tab:results-all}). The error bars mark the highest and the lowest bin values achieved in simulations with parameters set 1$\sigma$ off their respective medians in all possible combinations ($2^5 = 32$ combinations). \label{fig:bestplot-M4}}
\end{center}
\end{figure}

Among the models considered in this paper, the lowest $\chi^2$ values for the complete main sample were achieved for model M4 ($K=1$ and a double Gaussian radio loudness distribution). We plot the marginal probability distributions for the parameters of the best-fit model M4 in Figure \ref{fig:mcmarginal-M4}. Note that the expected scatter in $\chi^2$ values between simulations (due to a limited number of realizations) was taken into account prior to constructing the probability density distributions (see Figure~\ref{fig:nstack} and Section~\ref{sec:chi2evaluation}\ for details). The median values and 1$\sigma$ uncertainties derived from the marginal probability distributions for all five parameters of the model are given in Table~\ref{tab:results-all}. In Figure~\ref{fig:bestplot-M4}\ we plot the $t$, $i$, $z$ and $R'$ distributions for the observed SDSS-FIRST data (the main sample) and the median values obtained from a simulation (1000 individual realizations) performed with model M4 and the best-fit set of parameters. The error bars on each bin are the minimum and the maximum value occurring in that bin when parameters are shifted randomly within $\pm1\sigma$ from their respective best-fit medians.

\begin{figure}
\includegraphics[scale=0.5]{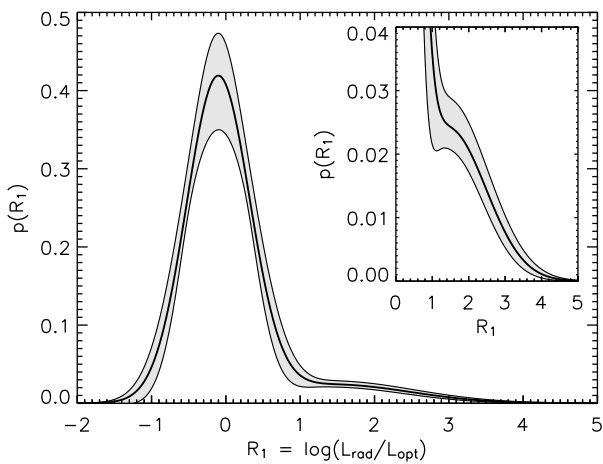}
\caption{Plot of the radio loudness distribution for median values of the M4 model parameters (thick black line). The shaded area marks propagated 1$\sigma$ uncertainties on the parameters. The inset shows a zoom-in on the region where the radio-loud regime starts to dominate. \label{fig:rifunction}}
\end{figure}

In Figure~\ref{fig:rifunction}\ we show the radio loudness distribution for our best-fit model M4, as well as the associated 1$\sigma$ uncertainties. We compute the probability of bimodality of the best-fit radio loudness distribution by deriving the fraction of the M4 model parameters which result in a bimodal distribution. We consider a distribution bimodal if there is at least a slight minimum between the RL and RQ regimes. We infer a probability of $\sim$20\% when we take the 1$\sigma$ confidence intervals of each parameter into account. Thus, we conclude that the likelihood of bimodality in the radio loudness distribution of our best-fit model is relatively small. Note, however, that based on our simulation tests (Section~\ref{sec:artif}) the total $\chi^2$ value of our best-fit model (1000 for 34 degrees of freedom) indicates that it is not a statistically acceptable representation of the data. In order to find a better-fitting model, in the next section we test for possible dependence of the M4 model parameters on redshift and optical luminosity.

\subsection{Dependence of the Best-fit Model Parameters on Redshift and Optical Luminosity}

In order to investigate whether a certain parametrization of dependence on redshift or optical luminosity could be added to our M4 model, we have divided our initial main sample into 4 bins in redshift and 4 bins in apparent magnitude and repeated the optimization procedure for each of them. Since most of these 16 subsamples span fairly narrow ranges in redshift and optical magnitude, the fits were done using only radio magnitude ($t$) and $R'=0.4(i-t)$ distributions and hence the number of degrees of freedom for this case is 14 per subsample. The choice for the binning was such that every subsample contains $\sim$6000 optical and $\sim$500 radio quasars, which is still large enough to perform fits to a distribution divided into 10 bins. The binning is given in Table~\ref{tab:izbins}, along with the results for each subsample.

Each of the 16 subsamples was best fitted with model M4, although for the optically faintest subsamples model M3 was almost equally well fit. $\chi^2$ values for individual subsamples range from 10 to 32 for 14 degrees of freedom and the lowest total $\chi^2$ value summed over all 16 subsamples is 291 for 224 degrees of freedom (reduced $\chi^2$ is 1.3). This is much lower than the lowest $\chi^2$ value for the M4 model fitted to the complete main sample, which is 1053 for 34 degrees of freedom. This result further confirms that a simple M4 model, which represents the distribution of the radio-to-optical ratio as a universal function independent of redshift and/or optical luminosity, is not a satisfactory representation of the real quasar sample from SDSS and FIRST.

\begin{figure}
\includegraphics[scale=0.67]{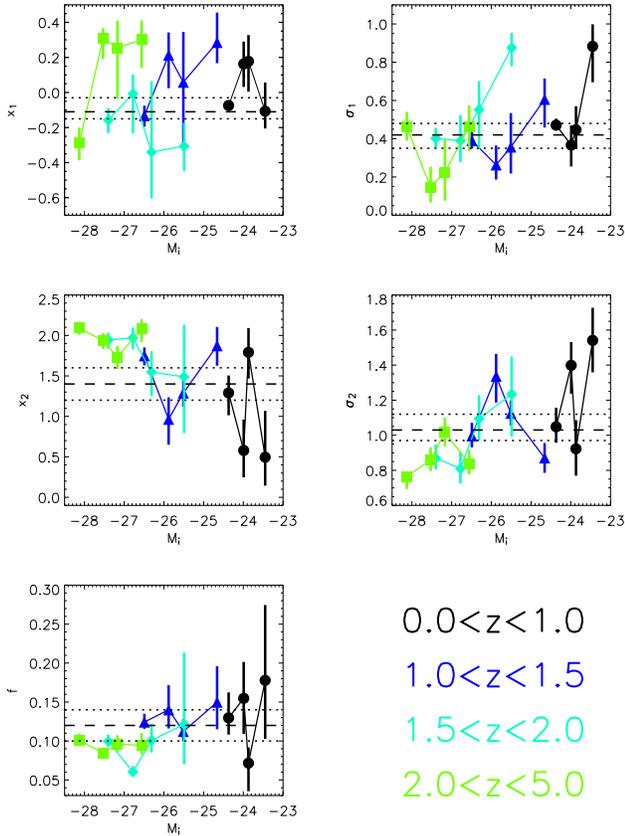}
\caption{ Each of the five panels in this figure shows the variation of best-fit parameters of the model M4 with absolute magnitude and redshift. Redshift and absolute magnitudes plotted are the medians of the subsamples (see Table~\ref{tab:izbins}). Colors and symbols mark different redshift bins. The dashed horizontal lines mark best-fit values for the complete main SDSS-FIRST sample and the dotted lines indicate 1$\sigma$ uncertainties. See Section~5.1 for a discussion of possible trends. \label{fig:izbins}}
\end{figure}

The results are plotted in Figure~\ref{fig:izbins} using median redshift and absolute optical magnitudes of the subsamples, showing that there are no clear trends in parameters $x_1$ and $\sigma_1$ and that possible trends exist in the remaining three parameters as a combination of dependence on redshift and/or optical luminosity. Results for the optically brightest subsample of each redshift bin were used to plot the continuous change of the radio loudness distribution in Figure~\ref{fig:shapeshift}. With the current results it is not possible to disentangle the two dependencies or to quantify them, but we do discuss the tentative trends in the following section.

\begin{table*}[!ht]
\caption{ The bin limits ($i_{\mbox{min}}$, $i_{\mbox{max}}$, $z_{\mbox{min}}$, $z_{\mbox{max}}$), median optical magnitudes ($i_{\mbox{med}}$ and $M_{i,\mbox{med}}$) and median redshifts ($z_{\mbox{med}}$) for the 16 subsamples used to examine changes in best-fit parameters of the M4 model with redshift and optical luminosity. Best-fit parameters of the M4 model and their associated 1$\sigma$ uncertainties are given for each of our 16 subsamples. \label{tab:izbins}}
\begin{center}
\begin{tabular}{|c|c|c|c|c|c|c|c|c|c|c|c|c|}
\hline
$i_{\mbox{\tiny MIN}}$ & $i_{\mbox{\tiny MED}}$ & $i_{\mbox{\tiny MAX}}$ & $z_{\mbox{\tiny MIN}}$ & $z_{\mbox{\tiny MED}}$ & $z_{\mbox{\tiny MAX}}$ & $M_{i,\mbox{\tiny MED}} $ & ${\chi^2}_{\mbox{\tiny MIN}}$ & $x_1$ & $\sigma_1$ & $f$ & $x_2$ & $\sigma_2$ \\
\hline \hline
15.0 & 18.06 & 18.5 & 0.0 & 0.53 & 1.0 & -24.37 & 27.9 & $ -0.07_{-0.03}^{+0.03} $ & $ 0.47_{-0.03}^{+0.03} $ & $ 0.13_{-0.02}^{+0.03} $ & $ 1.3_{-0.3}^{+0.2} $ & $ 1.0_{-0.1}^{+0.1} $ \\ 
18.5 & 18.78 & 19.0 & 0.0 & 0.61 & 1.0 & -23.99 & 15.9 & $ -0.2_{-0.1}^{+0.1} $ & $ 0.4_{-0.1}^{+0.1} $ & $ 0.15_{-0.05}^{+0.05} $ & $ 0.6_{-0.3}^{+0.4} $ & $ 1.4_{-0.2}^{+0.1} $ \\ 
19.0 & 19.13 & 19.5 & 0.0 & 0.66 & 1.0 & -23.87 & 19.7 & $ 0.2_{-0.2}^{+0.1} $ & $ 0.4_{-0.1}^{+0.1} $ & $ 0.07_{-0.04}^{+0.02} $ & $ 1.8_{-0.3}^{+0.3} $ & $ 0.9_{-0.2}^{+0.2} $ \\ 
19.5 & 19.94 & 21.5 & 0.0 & 0.77 & 1.0 & -23.45 & 14.1 & $ -0.1_{-0.1}^{+0.2} $ & $ 0.9_{-0.2}^{+0.1} $ & $ 0.17_{-0.08}^{+0.09} $ & $ 0.5_{-0.4}^{+0.6} $ & $ 1.5_{-0.2}^{+0.2} $ \\ 
\hline
15.0 & 18.16 & 18.5 & 1.0 & 1.23 & 1.5 & -26.49 & 29.4 & $ -0.13_{-0.06}^{+0.06} $ & $ 0.39_{-0.04}^{+0.05} $ & $ 0.12_{-0.01}^{+0.01} $ & $ 1.7_{-0.1}^{+0.1} $ & $ 1.00_{-0.07}^{+0.07} $ \\ 
18.5 & 18.79 & 19.0 & 1.0 & 1.24 & 1.5 & -25.88 & 9.5 & $ -0.2_{-0.2}^{+0.1} $ & $ 0.3_{-0.1}^{+0.1} $ & $ 0.14_{-0.02}^{+0.03} $ & $ 1.0_{-0.3}^{+0.3} $ & $ 1.3_{-0.2}^{+0.1} $ \\ 
19.0 & 19.22 & 19.5 & 1.0 & 1.26 & 1.5 & -25.51 & 9.6 & $ 0.1_{-0.4}^{+0.3} $ & $ 0.4_{-0.1}^{+0.2} $ & $ 0.11_{-0.01}^{+0.02} $ & $ 1.3_{-0.2}^{+0.1} $ & $ 1.13_{-0.07}^{+0.07} $ \\ 
19.5 & 20.05 & 21.5 & 1.0 & 1.26 & 1.5 & -24.66 & 17.1 & $ 0.3_{-0.1}^{+0.2} $ & $ 0.6_{-0.1}^{+0.1} $ & $ 0.15_{-0.03}^{+0.05} $ & $ 1.9_{-0.2}^{+0.2} $ & $ 0.87_{-0.09}^{+0.08} $ \\ 
\hline
15.0 & 18.18 & 18.5 & 1.5 & 1.74 & 2.0 & -27.40 & 32.1 & $ -0.16_{-0.07}^{+0.07} $ & $ 0.40_{-0.05}^{+0.05} $ & $ 0.10_{-0.01}^{+0.01} $ & $ 1.9_{-0.1}^{+0.1} $ & $ 0.87_{-0.06}^{+0.08} $ \\ 
18.5 & 18.79 & 19.0 & 1.5 & 1.73 & 2.0 & -26.78 & 10.8 & $ 0.0_{-0.2}^{+0.1} $ & $ 0.4_{-0.1}^{+0.1} $ & $ 0.06_{-0.01}^{+0.01} $ & $ 2.0_{-0.1}^{+0.1} $ & $ 0.81_{-0.08}^{+0.09} $ \\ 
19.0 & 19.25 & 19.5 & 1.5 & 1.72 & 2.0 & -26.31 & 30.2 & $ -0.3_{-0.3}^{+0.4} $ & $ 0.6_{-0.2}^{+0.1} $ & $ 0.10_{-0.02}^{+0.02} $ & $ 1.5_{-0.3}^{+0.3} $ & $ 1.1_{-0.1}^{+0.1} $ \\ 
19.5 & 20.09 & 21.5 & 1.5 & 1.74 & 2.0 & -25.49 & 14.4 & $ -0.3_{-0.1}^{+0.1} $ & $ 0.9_{-0.1}^{+0.1} $ & $ 0.12_{-0.05}^{+0.09} $ & $ 1.5_{-0.7}^{+0.6} $ & $ 1.2_{-0.2}^{+0.2} $ \\ 
\hline
15.0 & 18.19 & 18.5 & 2.0 & 2.30 & 5.0 & -28.13 & 16.5 & $ -0.3_{-0.1}^{+0.1} $ & $ 0.46_{-0.07}^{+0.08} $ & $ 0.10_{-0.01}^{+0.01} $ & $ 2.1_{-0.1}^{+0.1} $ & $ 0.76_{-0.07}^{+0.01} $ \\ 
18.5 & 18.80 & 19.0 & 2.0 & 2.30 & 5.0 & -27.53 & 27.0 & $ 0.3_{-0.1}^{+0.1} $ & $ 0.2_{-0.1}^{+0.1} $ & $ 0.08_{-0.01}^{+0.01} $ & $ 1.9_{-0.1}^{+0.1} $ & $ 0.85_{-0.06}^{+0.07} $ \\ 
19.0 & 19.20 & 19.5 & 2.0 & 2.34 & 5.0 & -27.17 & 9.6 & $ 0.3_{-0.3}^{+0.2} $ & $ 0.2_{-0.1}^{+0.2} $ & $ 0.10_{-0.01}^{+0.01} $ & $ 1.7_{-0.1}^{+0.1} $ & $ 1.02_{-0.08}^{+0.09} $ \\ 
19.5 & 20.04 & 21.5 & 2.0 & 2.98 & 5.0 & -26.96 & 10.1 & $ 0.3_{-0.2}^{+0.1} $ & $ 0.5_{-0.1}^{+0.1} $ & $ 0.09_{-0.01}^{+0.02} $ & $ 2.1_{-0.1}^{+0.2} $ & $ 0.84_{-0.06}^{+0.08} $ \\ 
\hline
\end{tabular}
\end{center}
\end{table*}

\section{Discussion}
\label{sec:discussion}

\subsection{Implications from the Parameters of the Best-fit Model}

Our best-fit model hints at the possibility that there might exist two distinct populations of quasars, assuming that the double-Gaussian parametrization is an appropriate one. Considering the fraction of radio loudness distributions that show bimodality within 1$\sigma$ confidence intervals of our best-fit model parameters, the probability that the global radio loudness distribution is bimodal is $\sim$20\%. Therefore, we can neither confirm nor firmly exclude that the radio loudness distribution is bimodal. A more robust result, however, is that the radio loudness distribution of SDSS-FIRST quasars can be described much better with two Gaussians than with a single one. This would imply that RL quasars obey a different relationship between radio and optical luminosity compared to RQ quasars. In this paper we parametrize our models so that there is either no relationship between radio and optical luminosities ($K=0$ models, M1 and M2), or $L_{\mbox{\tiny{1.4GHz}}} = L_{\mbox{\tiny{i-band}}}\times10^{R_1}$ ($K=1$ models, M3 and M4) for both types of quasars. The latter parametrization, for which we find a significantly better fit than for the former one, implies that the constant terms of the relationship (parameters $x_1$ and $x_2$, locations of the two Guassian peaks in $R_1$) and its scatter ($\sigma_1$ and $\sigma_2$, widths of the Gaussians in $R_1$) are different for the two classes of quasars. In general, the proportionality between the radio and optical luminosity might be different for each of the two quasar classes, but this type of a model was not considered in the work presented here.

Investigating a possible dependence on redshift or optical luminosity in our SDSS-FIRST sample, we have found a statistically good fit when the full sample is divided into 16 subsamples, each fitted independently (reduced $\chi^2$=291/224=1.3, summed over the 16 subsamples). The parameters describing the RQ Gaussian ($x_1$ and $\sigma_1$) were found not to vary significantly, which is not unexpected since this Gaussian is constrained only by its fall-off towards the RL regime. A possible trend with the absolute optical magnitude is suggested for the parameters describing the RL Gaussian ($x_2$ and $\sigma_2$) and the radio-loud fraction ($f$). The optically bright quasars appear to be better described by an RL Gaussian which is 'louder' (larger $x_2$), a smaller dispersion ($\sigma_2$) and a lower radio-loud fraction ($f$) than the optically faint quasars. There is a slight possibility that the trends are biased by the redshift-luminosity correlation inherent in all flux-limited surveys, but our method was designed specifically to avoid that kind of bias. Our tests performed on artificial datasets (see Section~\ref{sec:artif}) lend confidence that model parameters can be recovered correctly to within statistical uncertainties under a variety of different conditions.

\begin{figure}
\includegraphics[scale=0.53]{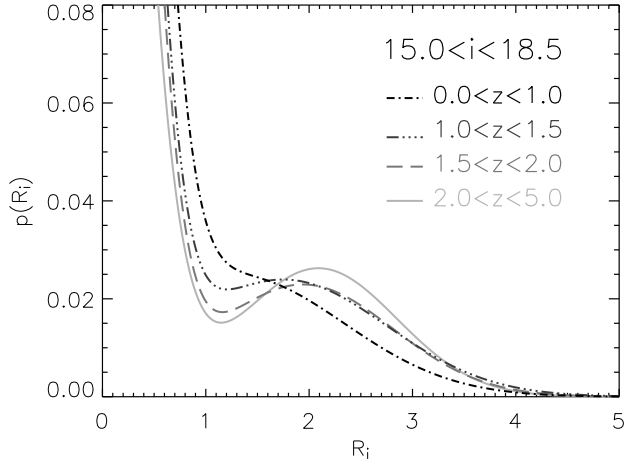}
\caption{ Radio loudness distribution as a function of redshift for flux-limited subsambles of quasars with $i<18.5$. Note that bimodality seems to have been more prominent at high redshift and that it flattened out at low redshift. The same general trend is detectable in fainter subsamples, but the shape of the distribution does not change in a monotonous manner. \label{fig:shapeshift}}
\end{figure}

Figure~\ref{fig:shapeshift} shows a possibly significant change of the radio loudness distribution shape implied from our results for quasars with $i<18.5$ (the brightest subsample in each redshift bin). This apparent shift of the RL peak and the change in its width could be indicative that the radio loudness distribution becomes increasingly bimodal at high redshift. A monotonous change in the radio loudness distribution shape is clearly visible for our subsamples with $i<18.5$, indicating that radio-loud quasars are rarer, 'louder' and less spread out in their loudness at high redshift. The same general trend is roughly detectable for subsamples with $i>18.5$, but the change of the radio loudness distribution shape is not nearly as clear and continuous as for the $i<18.5$ subsamples. We wish to emphasize that we understand selection effects for the $i<18.5$ subsample much better than for the $i>18.5$ subsample. First, the quasar targeting is complete only for $i<19$ candidates. Second, the quasar variability will introduce an RMS scatter of 0.2-0.3 magnitudes and thus blur this supposedly sharp flux limit. Hence, our choice of $i<18.5$ binning defines the faintest sample for which the simple SDSS optical selection function is highly reliable. The lack of clear and continuous change in fainter samples may be reflecting the lower reliability of the $i>18.5$ objects in the SDSS quasar sample, or indicate that the dependence of the radio loudness distribution on redshift or optical luminosity is not monotonous. We plan to investigate this in the future with more complex model and different statistics (e.g.\ maximum likelihood), so that more information may be extracted from the existing data.

\subsection{Comparison to Previous Findings}

Previous work on the issue of the bimodality in the radio loudness distribution of quasars were either based on small samples of quasars, or large but less reliable ones. For example, \citet{white00} and \citet{cirasuolo03a,cirasuolo03b} used spectroscopically confirmed samples of 636, 141 and 195 radio-detected quasars, respectively, while \citet{ivezic02,ivezic04} used photometric samples with $\sim$4400 and $\sim$10,000 radio-detected quasar candidate sources. In comparison to those results, the results presented here have considerably better statistics and reliability. Our result that a two-component model with a direct proportionality between radio and optical luminosities fits the observed data best is consistent with \citet{cirasuolo03b}, who have used a method similar to ours on a much smaller heterogeneous sample. Using a stacking analysis of the FIRST data to probe faint radio fluxes, \citet{white07} also found a strong dependence of the median radio luminosity of quasars on the optical luminosity.

\begin{figure}
\includegraphics[scale=0.71]{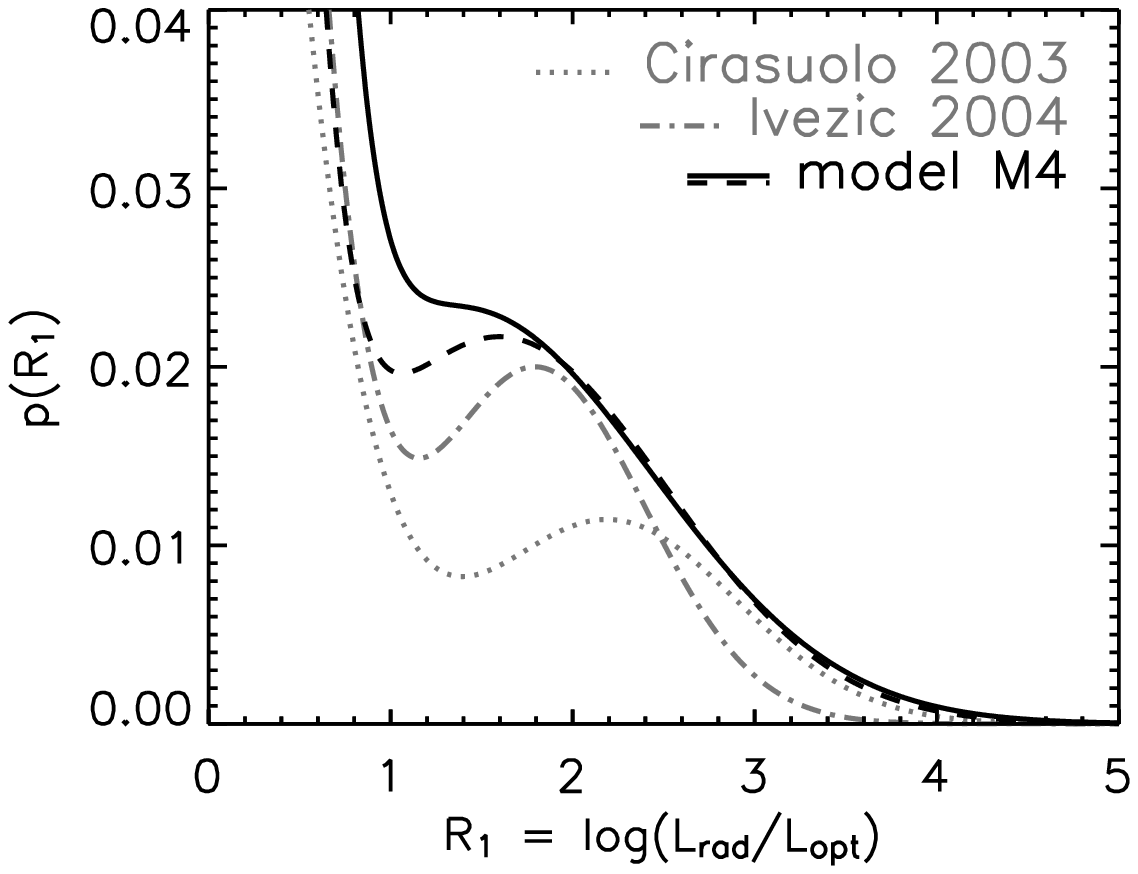}
\caption{ A comparison of our results (thick solid and dashed lines) to the previous results by \citet{cirasuolo03b} and \citet{ivezic04} (dotted and dot-dashed grey lines, respectively). The results from \citet{cirasuolo03b} were shifted by 0.4 towards lower values to account for different optical bands, $i$ versus $B$. The thick solid line is the intrinsic radio loudness distribution ($R_1$) with parameters given in Table~\ref{tab:results-all} for model M4. The thick dashed line shows the result one would obtain with our sample by not accounting for the incompleteness of FIRST survey near its flux limit. \label{fig:usporedba}}
\end{figure}

Our results can be compared to some of the previous ones as plotted in Figure~\ref{fig:usporedba}. The prominent minimum separating the RQ and RL Gaussians observed at $R_1\sim$1.3 by \citet{cirasuolo03b} and \citet{ivezic04} is likely due to the incompleteness of the FIRST survey at its faint end. In the region around $R_1\sim1$, FIRST is not more than $\sim$70\% complete for the faintest radio sources, but this was not taken into account in the earlier works. The plot also shows how our result would change in the case where we neglect FIRST incompleteness and assume it is 100\% complete down to its flux limit -- in that case we would have found a weak bimodality with a minimum at $R_1\sim$1. Its relative weakness in comparison to previous results can be understood as a difference in the fraction of optically faint quasars in the sample, which is lower in our case (45\% here compared to $\sim$75 in other cases), and which limits the influence of incompleteness to lower values of $R_1$. Note that the flux limit in the radio is the same for all three studies compared in Figure~\ref{fig:usporedba}, so optically fainter samples introduce incompleteness at a higher value of $R_1$. Some of the difference with previous results may also be related to the uncertainties arrising from K-corrections and optical variability.
 
A small degree of discrepancy exists in the fraction of RL quasars between our and previous results; $f=(12\pm1)\%$ (inferred here) compared to $(8\pm1)\%$ from \citet{ivezic04} and $(3\pm2)\%$ from \citet{cirasuolo03b}. We have defined the radio-loud quasar fraction as the ratio between the areas under the RL and RQ Gaussians, denoted in our models as parameter $f$. In other words, the fraction of RL quasars is the fraction of quasars whose radio loudness is determined by the RL Gaussian. Note that with this definition the range of radio loudness between the RL and RQ peaks is shared by both types of quasars. \cite{cirasuolo03b} have used that definition as well, but \cite{ivezic02,ivezic04} defined radio-loud quasars as the ones having $R_1>1$. This definition is more practical and a radio loudness may be unambiguously assigned to each individual quasar. With this definition, our RL fraction is $(10\pm1)\%$, consistent with the results of \citet{ivezic04}.

\begin{figure}
\includegraphics[scale=0.51]{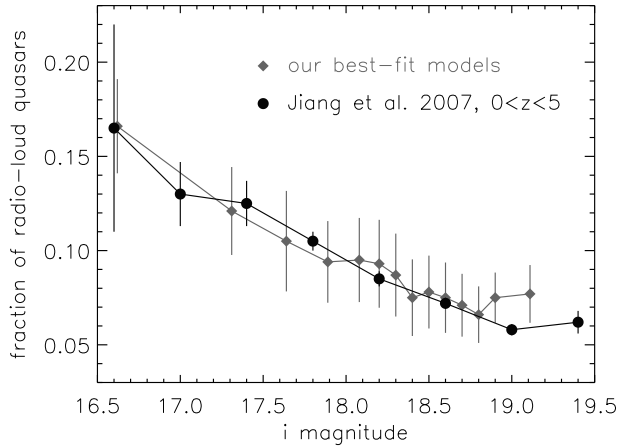}
\caption{ Dependence of the radio-loud fraction, defined with a threshold in radio loudness and calculated from a best-fit model, as a function of apparent $i$ magnitude (grey symbols and lines). For comparison, the same result from Jiang et~al.~(2007) is plotted in black. M4 model parameters for each subsample were obtained independently and radio-loud fraction was calculated by integrating the analytic form of the best-fit radio loudness distribution function (e.g.\ it is not the $f$ parameter of model M4). \label{fig:narbins}}
\end{figure}

\citet{jiang07} have found that the fraction of RL quasars depends both on optical luminosity and redshift, practically independent of the exact definition of an RL quasar. They have found that the fraction of RL quasars, when defined with a threshold in radio loudness is higher for optically bright quasars and at low redshift. Our results tentatively confirm that such trends exist, albeit at low significance (see the lowest panel in Figure~\ref{fig:izbins}). With our definition of the RL fraction (the fraction of quasars in the radio-loud Gaussian, $f$) the trend with respect to the optical luminosity seems to be just the opposite -- $f$ tends to be lower for optically brighter quasars. Note, however, that even if the radio loudness distribution shape changes (e.g.\ as in our $i<18.5$ results shown in Figure~\ref{fig:shapeshift}), the fraction of RL quasars defined with a threshold in radio loudness can remain constant or have an opposite trend than parameter $f$. In order to check for consistency with \citet{jiang07}, we have performed additional fits to subsamples derived from the original main sample by dividing it into narrow bins in apparent optical magnitude. As displayed in Figure~\ref{fig:narbins}, the fraction of RL quasars calculated from the best-fit M4 models for each subsample, using a threshold in radio loudness, matches the data from \citet{jiang07} very well.

In broad agreement with previous work by other authors, our current results imply that the radio loudness distribution did not have the same shape at all times in cosmic history. It might be possible to investigate this in the future using deep radio imaging of samples in thin redshift slices. For example, \citet{kimball11} have used deep imaging with EVLA to probe a statistically complete sample of quasars in a narrow redshift bin at $0.2<z<0.3$. They found that the radio loudness distribution can be well explained with a superposition of two QSO populations (radio-loud, which is AGN-dominated and radio-quiet, dominated by star formation in the host galaxy), and their distribution of radio luminosity appears similar to the best-fit two-component model presented here. Since currently available survey data lacks either depth in the radio or significant sky coverage, the data from the newer generation of radio instrumentation (JVLA, SKA) will be essential in order to fully constrain the radio loudness distribution and ultimately, understand its physical origin.

\section{Summary}
\label{sec:summary}

In this paper we present Monte Carlo simulations of the optically-selected quasar population fine-tuned to study the radio loudness distribution of quasars. We investigate four different models based on single and double Gaussian distributions and different radio-to-optical luminosity relationships. Our aim was to investigate the long-standing ambiguity in the existence of the intrinsic difference between radio-quiet and radio-loud quasars and in particular, the existence of bimodality in their radio loudness distribution. The sample used here, based on SDSS DR7 Quasar Catalog matched to the FIRST survey, is the largest ever analyzed (8307 radio-detected quasars), uniformly selected and reliable (spectroscopically confirmed), and the method properly accounts for uncertainties in the K-corrections, optical variability and survey incompleteness. 

We find that the best-fit model for this SDSS-FIRST quasar sample is a two-component model, but the components overlap so that radio loudness bimodality (i.e.\ a minimum between the radio loud and quiet portions of the distribution) is not apparent. Statistics of our fits indicate that even our best-fit model does not describe the data optimally, although it is significantly better than any other model we used. Our main result, that the radio loudness distribution of quasars likely consists of at least two components, agrees with earlier findings indicating the existence of two distinct populations of quasars. In the framework of the simple two-component models presented here, we conclude that bimodality is not likely in a sample which, like ours, covers a broad redshift and optical luminosity range, even if it is present in narrower redshift and magnitude bins (e.g. at $z>1.5$ and $i<18.5$). We also conclude that the distribution is unlikely to be universal, i.e.\ independent of redshift or optical luminosity.

Investigating possible redshift and optical luminosity dependence of the radio loudness distribution, we have found that a monotonous change of its shape is visible for quasars with $i<18.5$. It would imply that at high redshift radio-loud quasars were rarer, on average 'louder' and less spread out in radio loudness. The same general trend is marginally detectable for fainter quasars in the sample, but the smooth change of the radio loudness distribution shape is not nearly as apparent as for $i<18.5$. We further conclude that the radio loudness distribution is likely dependent on redshift and/or optical luminosity, but we can not disentangle the two dependencies with current models and data. We expect more sophisticated two-component models to adequately describe the two classes of quasars in future work on this topic.

The research leading to these results has received funding from the European Union's Seventh Framework programme under grant agreement 229517. \v{Z}.~I. acknowledges support by NSF grant AST-0807500 to the University of Washington, NSF grant AST-0551161 to LSST for design and development activity, and by the Croatian National Science Foundation grant O-1548-2009. V.~S. acknowledges support from NASA grant HST-GO-09822.31-A. M.~B. acknowledges support from the International Fulbright Science and Technology Award.

The National Radio Astronomy Observatory is a facility of the National Science Foundation operated under cooperative agreement by Associated Universities, Inc. Funding for the SDSS and SDSS-II has been provided by the Alfred P. Sloan Foundation, the Participating Institutions, the National Science Foundation, the U.S. Department of Energy, the National Aeronautics and Space Administration, the Japanese Monbukagakusho, the Max Planck Society, and the Higher Education Funding Council for England. The SDSS Web Site is http://www.sdss.org/. The SDSS is managed by the Astrophysical Research Consortium for the Participating Institutions. The Participating Institutions are the American Museum of Natural History, Astrophysical Institute Potsdam, University of Basel, University of Cambridge, Case Western Reserve University, University of Chicago, Drexel University, Fermilab, the Institute for Advanced Study, the Japan Participation Group, Johns Hopkins University, the Joint Institute for Nuclear Astrophysics, the Kavli Institute for Particle Astrophysics and Cosmology, the Korean Scientist Group, the Chinese Academy of Sciences (LAMOST), Los Alamos National Laboratory, the Max-Planck-Institute for Astronomy (MPIA), the Max-Planck-Institute for Astrophysics (MPA), New Mexico State University, Ohio State University, University of Pittsburgh, University of Portsmouth, Princeton University, the United States Naval Observatory, and the University of Washington.


{}

\end{document}